\documentclass[prl,twocolumn,showpacs]{revtex4-1}

\newcommand{\deltazmax}{(\delta z)_\mathrm{max}}
\newcommand{\Eq}[1]{Eq.~(\ref{eq:#1})}

\newcommand{\Fig}[1]{Fig.~\ref{fig:#1}}

\newcommand{\Ref}[1]{Ref.~\cite{#1}}

\newcommand{\f}{\mathbf{f}}
\newcommand{\gdot}{\dot\gamma}
\newcommand{\gdotmax}{\gdot_\mathrm{max}}
\newcommand{\rr}{\mathbf{r}}
\renewcommand{\v}{\mathbf{v}}

\usepackage{graphicx}

\begin{document}
\title{Dimensionality and viscosity exponent in shear-driven jamming}

\author{Peter Olsson}
\affiliation{Department of Physics, Ume\aa\ University, 901 87 Ume\aa, Sweden}

\date{\today}   
\begin{abstract}
  Collections of bidisperse frictionless particles at zero temperature in three dimensions
  are simulated with a shear-driven dynamics with the aim to compare with behavior in two
  dimensions. Contrary to the prevailing picture, and in contrast to results from
  isotropic jamming from compression or quench, we find that the critical exponents in
  three dimensions are different from those in two dimensions and conclude that
  shear-driven jamming in two and three dimensions belong to different universality
  classes.
\end{abstract}

\maketitle

\paragraph{Introduction}

A system of granular particles at zero temperature with contact-only interactions,
undergoes a jamming transition, which is a transition from a liquid to a disordered solid,
at a critical packing fraction $\phi_J$. As this is a phenomenon at zero temperature,
there is no thermal equilibrium and it turns out that details of the jamming transition
depend on the physical protocol by which the system jams; \emph{isotropic} jamming and
\emph{shear-driven} jamming thus appear to be different phenomena.

Isotropic jamming results when the system is either compressed isotropically
\cite{Donev_TS:packing, Berthier_Witten:PRE2009, Vagberg_OT:protocol, Ozawa:2017} or when
it is rapidly quenched from $T=\infty$ to $T=0$ at fixed volume
\cite{OHern_Langer_Liu_Nagel:2002, OHern_Silbert_Liu_Nagel:2003, Vagberg_OT:protocol}.  In
both cases the resulting jammed state has (in principle) an isotropic stress tensor. When
compressed the particle packing $\phi$ is increased by slowly and isotropically
compressing a system.  As $\phi$ increases, particles come into contact with each other,
at $\phi_J$ a mechanically stable rigid backbone of particles percolates across the
system, and the system jams.  The precise value of $\phi_J$ varies somewhat with the
details of the protocol for compressing or quenching \cite{Chaudhuri_Berthier_Sastry,
  Vagberg_OT:protocol}, as properties of the starting configurations and the rate of
compression or quench.

In shear-driven jamming of frictionless particles the system is sheared at constant volume
with a uniform shear strain rate $\dot\gamma$.  Below $\phi_J$---the jamming density of
the shear-driven jamming transition, which is independent of the initial configuration---
the system behaves as a liquid with a finite viscosity, $\lim_{\dot\gamma\to 0}
(\sigma/\dot\gamma)$, where $\sigma$ is the shear stress.  Above $\phi_J$ a finite yield
stress developes, $\lim_{\dot\gamma\to 0}\sigma>0$.

Early numerical simulations in 2D and 3D led to the conclusion that the critical exponents
associated with isotropic jamming are independent of the dimensionality of the
system \cite{OHern_Silbert_Liu_Nagel:2003}.  More recently it has been demonstrated
numerically that key non-trivial critical exponents for isotropic jamming agree
quite well \cite{Lerner-SoftMatter:2013, DeGiuli:PNAS:2014, Charbonneau:2015:prl} with the
values predicted analytically from an infinite-dimensional mean-field theory
\cite{Charbonneau:NatCommun:2014, Charbonneau:Jstat:2014}.  This observation has supported
earlier claims that the upper critical dimension for isotropic jamming is $d_u=2$
\cite{Wyart:2005, Goodrich:2012}, and that mean-field results apply for any $d>d_u$.  The
prevailing view has been that the same should be true for shear-driven jamming
\cite{DeGiuli:2015}, and theoretical models have been constructed that try to relate the
critical exponents for shear-driven jamming to the mean-field values appropriate to
isotropic jamming \cite{DeGiuli:2015, During_Lerner_Wyart:2016}.  In this work we
argue that this prevailing view is incorrect.  By extensive numerical simulations, and a
carefully quantitative analysis of the critical behavior, we show that the exponent
associated with the diverging viscosity below $\phi_J$ is clearly different in 2D and 3D,
thus demonstrating that shear-driven jamming in physical dimensions cannot be considered a
mean-field transition.

The expectation that shear-driven jamming in two and three dimensions should behave the
same, seems to be taken over from the above-mentioned dimension-independence found for
isotropic jamming, together with the common result that weakly driven systems may be
considered to be small perturbations about configurations in the absence of driving. This
is however not applicable in the present situation since the shearing may never be
considered to be a small perturbation. One way to see this is by considering the
dimensionless friction at criticality which is $\mu\equiv\sigma/p\approx 0.1$
\cite{Lerner-PNAS:2012} (where $p$ is pressure), which means that the system is highly
anisotropic even in the limit of weak driving. A situation when linear response \emph{is}
applicable is in shearing simulations at \emph{finite} temperatures and small $\gdot/T$
\cite{Olsson_Teitel:jam-T}, but it is then found that linear response is applicable only
as long as the system is close to isotropic, $\sigma/p<0.01$.


Several attempts have been made to determine the critical behavior of shear-driven jamming
\cite{Olsson_Teitel:jamming, Hatano:2008, Heussinger_Barrat:2009, Otsuki_Hayakawa:2009b,
  Hatano:2010, Tighe_WRvSvH, Olsson_Teitel:gdot-scale, Lerner-PNAS:2012, DeGiuli:2015,
  Kawasaki_Berthier:2015}. We here briefly review a few these methods. The first is to
determine shear stress, $\sigma(\phi,\gdot)$ or pressure, $p(\phi,\gdot)$ from
shear-driven simulations of soft disks at different densities and shear strain rates and
make use of a scaling relation, described below \cite{Olsson_Teitel:jamming,
  Olsson_Teitel:gdot-scale}, to try to extract the behavior in the $\gdot\to0$ limit. With
the pressure-equivalent of the shear viscosity, $\eta_p\sim p/\gdot$, the divergence at
the jamming density $\phi_J$ is governed by the exponent $\beta$,
\begin{equation}
  \label{eq:beta}
  \eta_p(\phi,\gdot\to0) \sim (\phi_J-\phi)^{-\beta}.
\end{equation}
Since the particle overlaps get smaller for smaller $\gdot$, the limit $\gdot\to0$ is the
hard particle limit.

Another way to approach criticality is by doing shearing simulations with hard particles
\cite{Lerner-PNAS:2012}. Since hard frictionless particles jam when the contact number is
equal to $z=z_c=2d$ (when $z$ is determined after removing the rattlers from the system)
the idea is to determine how $\eta_p$ diverges as $z_c$ is approached. With $z_c-z \sim
(\phi_J-\phi)^{u_z}$ \Eq{beta} becomes 
\begin{equation}
  \label{eq:betauz}
  \eta_p \sim (z_c - z)^{-\beta/u_z}.
\end{equation}
The advantage of this expression over \Eq{beta} is that $z_c$ is known whereas $\phi_J$ in
\Eq{beta} is unknown and has to be determined from the fitting together with the
exponent. \Eq{betauz} therefore opens up a more direct analysis by just plotting $\eta_p$
vs $\delta z\equiv z_c-z$.

\begin{figure}
  \includegraphics[width=8cm]{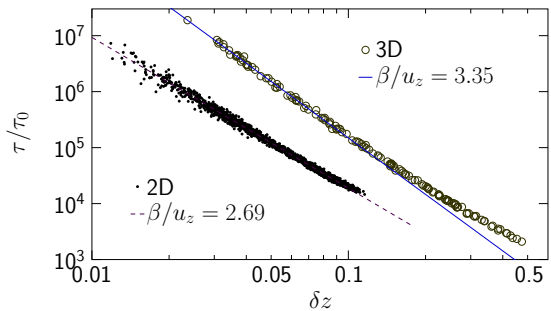}
  \caption{Relaxation time vs distance to the transition as measured by $\delta z\equiv
    z_c-z$. The figure shows results for both 2D and 3D and gives strong evidence that
    the exponents $\beta/u_z$, given by the slopes at small $\delta z$ in 3D and 2D, are
    different.}
  \label{fig:tau-dz}
\end{figure}

For comparing determinations of $\beta$ and $\beta/u_z$ one needs a value for $u_z$, which
in \Ref{Heussinger_Barrat:2009} was found to be $u_z=1$. This determination was however
done without first removing the rattlers, and the precision has also been questioned
\cite{DeGiuli:2015}. Turning things the other way around, $u_z$ in 2D may be determined
from $\beta= 2.70\pm0.15$ from the scaling analysis \cite{Olsson_Teitel:gdot-scale,
  beta2d} and $\beta/u_z= 2.69\pm0.03$ \cite{Olsson:jam-tau} (also shown in \Fig{tau-dz})
which gives $u_z=0.996\pm0.057$. Here and throughout the paper the quoted errors are max/min
values, three standard deviations.

The essence of the shear-driven jamming transition is the slowing down of the dynamics,
and the characterization of this dynamics is the idea behind a different but related
method to study the jamming transition. In this method the ordinary shearing at a fixed
$\gdot$ is suddenly stopped and the system is made to relax to vanishing energy
\cite{Olsson:jam-tau}. From the exponential relaxations one determines the relaxation time
$\tau$ while one measures $\delta z$ from the final configuration. (Note that this
relaxation time is not the same as the relaxation time, commonly determined in steady
state or at equilibrium, which is obtained from the self-part of the intermediate
scattering function \cite{Brambilla:2009}.)  It turns out that $\tau$ determined from such
relaxations behaves the same as $\eta_p$\cite{Olsson:jam-tau, Lerner-PNAS:2012} and we
have
\begin{equation}
  \label{eq:tau}
  \tau \sim (\delta z)^{-\beta/u_z}.
\end{equation}
The present paper presents shearing simulations of soft elastic particles and makes use of
both the scaling analysis of pressure and the analysis of the relaxation time.

\paragraph{Models and simulations}

For the simulations we follow O'Hern \emph{et al.}\cite{OHern_Silbert_Liu_Nagel:2003} and
use a simple model of bi-disperse frictionless soft particles---disks or balls---in two
and three dimensions with equal numbers of particles with two different radii in the ratio
1.4. Length is measured in units of the diameter of the small particles, $d_s$. We use
Lees-Edwards boundary conditions\cite{Evans_Morriss} to introduce a time-dependent shear
strain $\gamma = t\gdot$. We define the non-affine velocity, $\v_i = \dot\rr_i-
\v^\mathrm{aff}(\rr_i)$, obtained by subtracting off the uniform shear velocity
$\v^\mathrm{aff}(\rr_i)\equiv\gdot y_i\hat x$ from the particle center of mass velocity
$\dot\rr_i$. With $r_{ij}$ the distance between the centers of two particles and $d_{ij}$
the sum of their radii, the relative overlap is $\delta_{ij} = 1 - r_{ij}/d_{ij}$ and the
interaction between overlapping particles is $V(r_{ij}) = \epsilon \delta_{ij}^2/2$; we
take $\epsilon=1$. The force on particle $i$ from particle $j$ is $\f^\mathrm{el}_{ij} =
-\nabla_i V(r_{ij})$. The simulations are performed at zero temperature.

We consider the interaction force $\f^\mathrm{el}_i = \sum_j \f^\mathrm{el}_{ij}$ where
the sum extends over all particles $j$ in contact with $i$. The simulations discussed here
have been done with the RD$_0$ (reservoir dissipation) model with the dissipating force
$\f^\mathrm{dis}_i = -k_d \v_i$ \cite{Vagberg_Olsson_Teitel:BagnNewt}.  In the overdamped
limit the equation of motion is $\f^\mathrm{el}_i +\f^\mathrm{dis}_i = 0$ which becomes
$\v_i = \f^\mathrm{el}_i/k_d$.  We take $k_d=1/2$ and the time unit $\tau_0 =
d_s^2 k_d/\epsilon=1/2$. The equations of motion were integrated with the Heuns method
with time step $\Delta t/\tau_0=0.4$.  We simulate with $N=65536$ particles and shear
strain rates down to $\gdot\tau_0=10^{-8}$. By determining the $\gdot$ below which finite
size effects start become visible in additional simulations with $N=1024$ and $N=4096$, we
conclude that our data with $N=65536$ should not be affected by finite size effects.


Beside the ordinary simulations at constant shear strain rate we do two-step simulations
where the constant shearing is suddenly stopped but the dynamics is continued, such that
the systems relax to vanishing energy. From the exponential relaxations of $p$ we
determine the relaxation times $\tau$ and from the final configurations we determine the
contact number $z$, after first removing the rattlers. The values of $\tau$ and $\delta
z\equiv z_c-z$ from these relaxations will spread around averages that depend on both
$\phi$ and the initial $\gdot$. It does however turn out when plotting points
parametrically as $\tau$ vs $\delta z$, the points fall on a single common curve
independent of the starting parameters \cite{Olsson:jam-tau}. This observation may be
rationalized by considering that the final steps of the relaxation is probing the hard
particle limit in which the dynamics is controlled by the properties of the contact
network only, and thereby by the distance to jamming as measured by $\delta z$.

\paragraph{Results}

Our key result is summarized by \Fig{tau-dz} where $\tau$ vs $\delta z$ in both two and
three dimensions are shown by solid dots and open circles, respectively. The 2D data are
from \Ref{Olsson:jam-tau}. As always in the determination of critical exponents, we are
interested in the data closest to criticality, i.e.\ at small $\delta z$; we note that the
slopes at small $\delta z$ in \Fig{tau-dz} are clearly different.  Fitting data with
$\delta z<0.08$ to \Eq{betauz} gives the exponent $\beta/u_z=3.35\pm0.12$ in 3D, clearly
different from $\beta/u_z=2.69\pm0.03$ in 2D \cite{Olsson:jam-tau}. This is therefore
strong evidence that shear-driven jamming in 3D and 2D belong to different universality
classes.  A more careful determination of the 3D exponent is given below.

\begin{figure}
  \includegraphics[width=8cm]{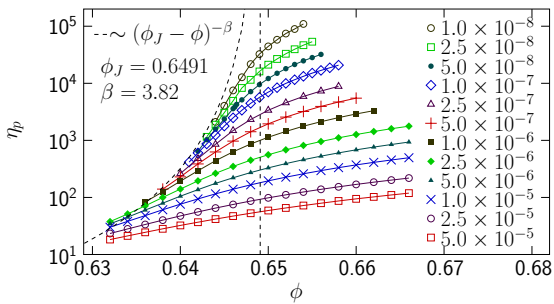}
  \caption{Values of $\eta_p$ used in the scaling analyses. Different curves are different
    shear strain rates. These data are for parameters that obey the conditions
    $0.632\leq\phi\leq0.666$ and $|X|<0.2$, where $X=(\phi-0.6491)/ \gdot^{0.205}$.}
  \label{fig:etap-phi}
\end{figure}

\paragraph{Scaling analysis}
For a more detailed characterization of the critical behavior we turn to a scaling
analysis of $p(\phi,\gdot)$. Following \Ref{Olsson_Teitel:gdot-scale} the starting point is the
scaling assumption below, where the second term gives the leading corrections-to-scaling,
\begin{equation}
  \label{eq:p-b}
  p(\delta\phi,\gdot) = 
  b^{-y/\nu} \left[ f(\delta\phi b^{1/\nu}, \gdot b^z) + 
    b^{-\omega} g(\delta\phi b^{1/\nu}, \gdot b^z)\right].
\end{equation}
Here $b$ is a length rescaling factor, $f$ and $g$ are scaling functions, $\nu$ is the
correlation length exponent, $z$ is the dynamical critical exponent, $y$ is the scaling
dimension of $p$, and $\omega$ is the correction-to-scaling exponent. Choosing $b$ so that
$\gdot b^z=1$, and with $q=y/z\nu$, this becomes
\begin{equation}
  \label{eq:p-scale}
  p(\delta\phi,\gdot) = \gdot^q \left[ f_p\left(\frac{\delta\phi}{\gdot^{1/z\nu}}\right)
  + \gdot^{\omega/z} g_p\left(\frac{\delta\phi}{\gdot^{1/z\nu}}\right) \right].
\end{equation}
We take $f_p$ and $g_p$ to be exponentials of sixth and third order polynomials in
$\delta\phi/\gdot^{1/z\nu}$, respectively. The data used for the fits are shown in
\Fig{etap-phi} as $\eta_p(\phi,\gdot)\equiv p/(\gdot\tau_0)$ for shear rates
$\gdot\tau_0=10^{-8}$ through $5\times10^{-5}$.

\begin{figure}
  \includegraphics[bb=31 324 370 580, width=4.2cm]{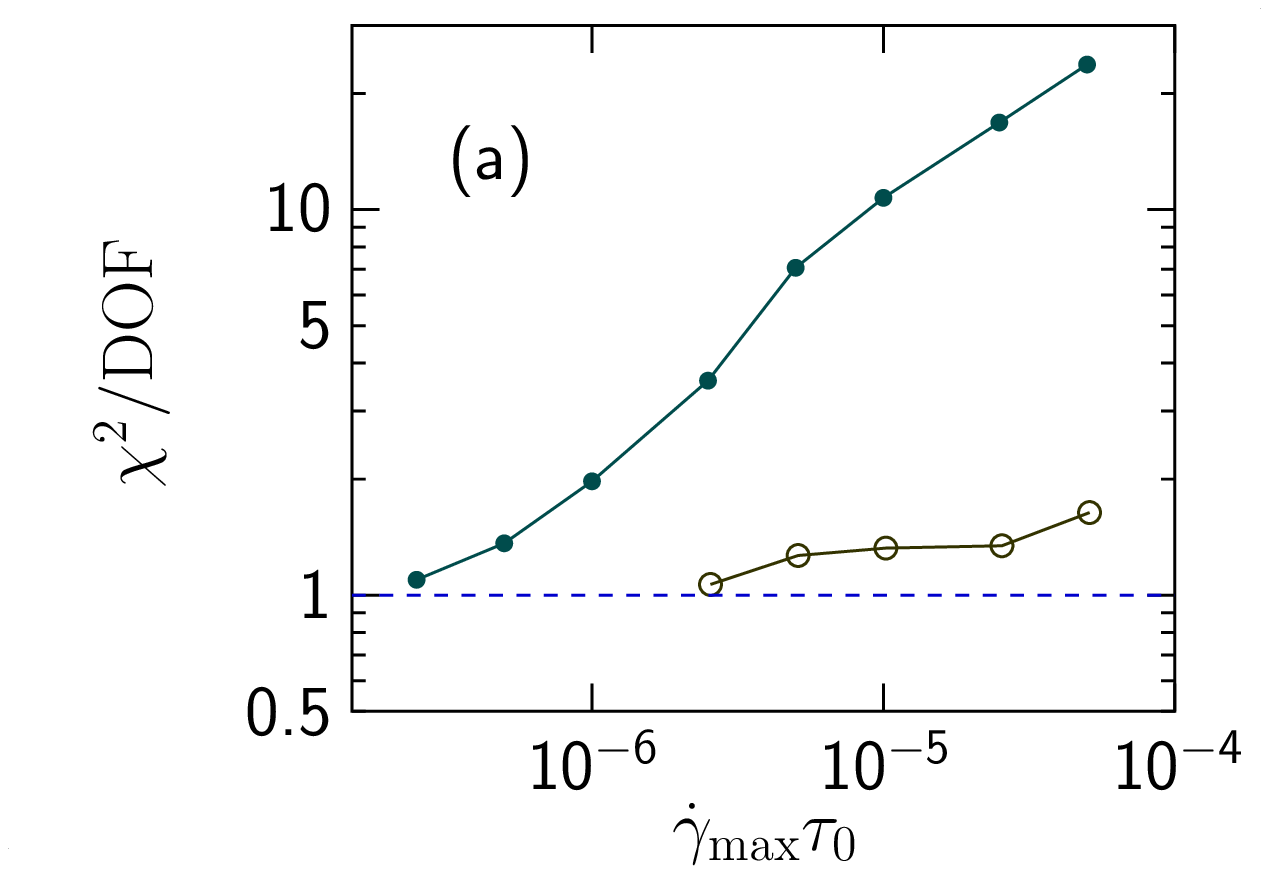}
  \includegraphics[bb=31 324 370 580, width=4.2cm]{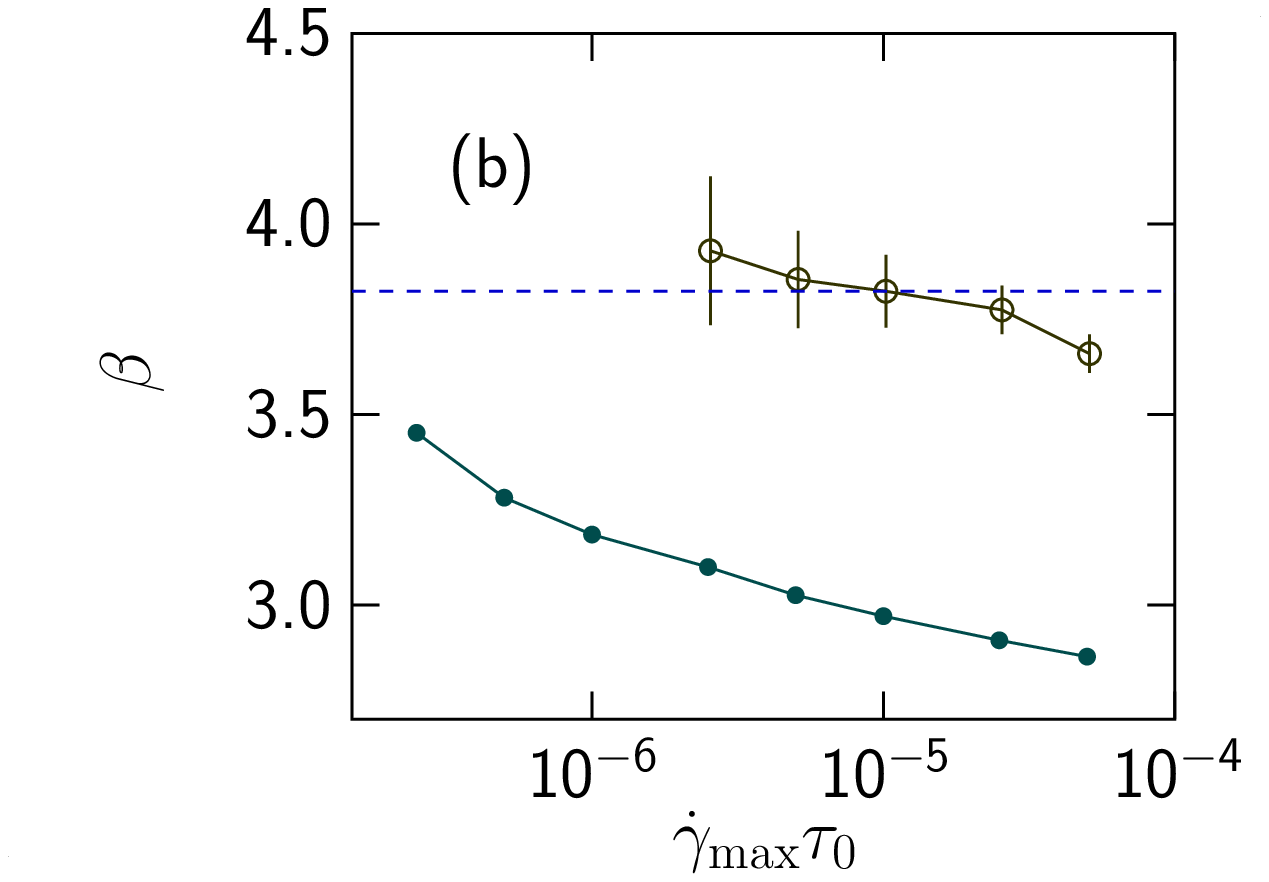}
  \includegraphics[bb=31 324 370 580, width=4.2cm]{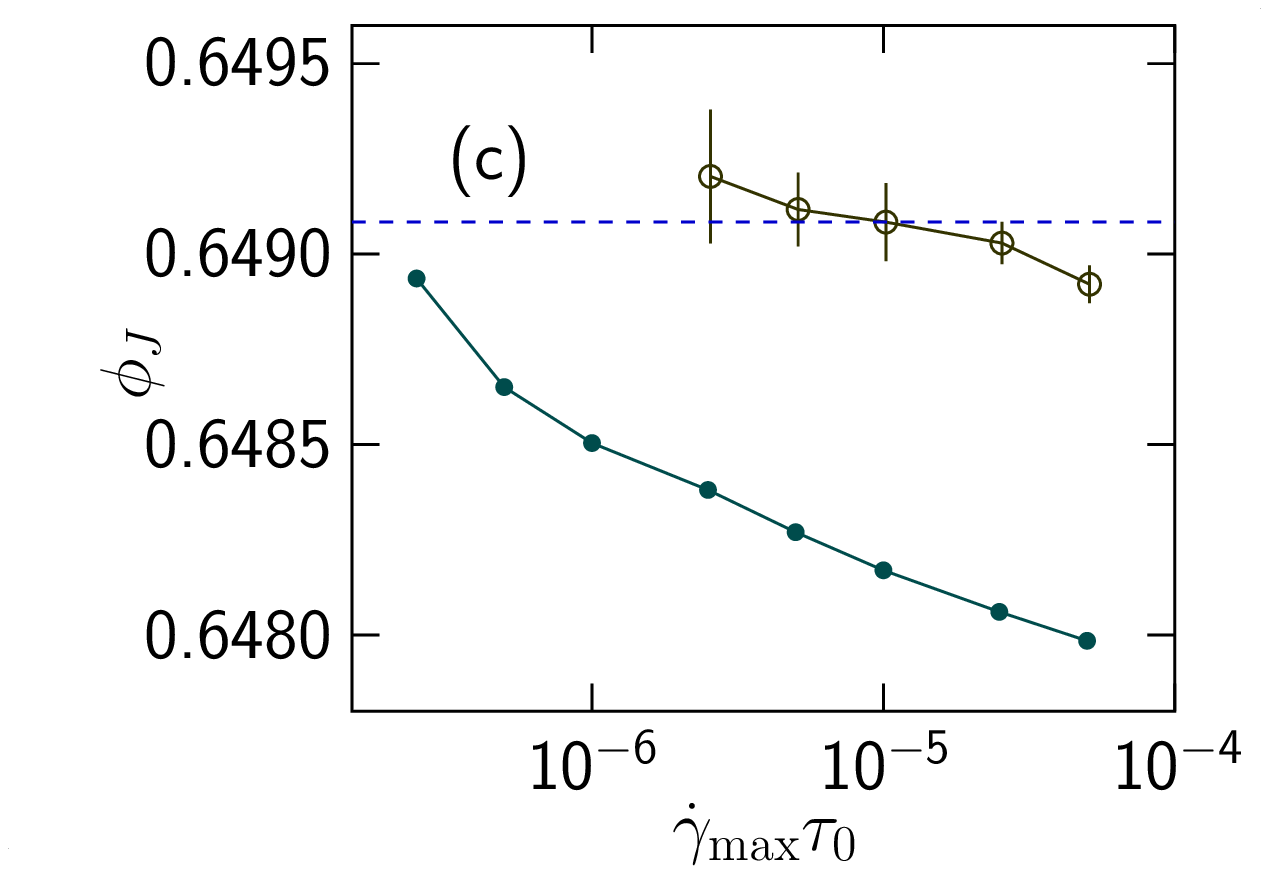}
  \includegraphics[bb=31 324 370 580, width=4.2cm]{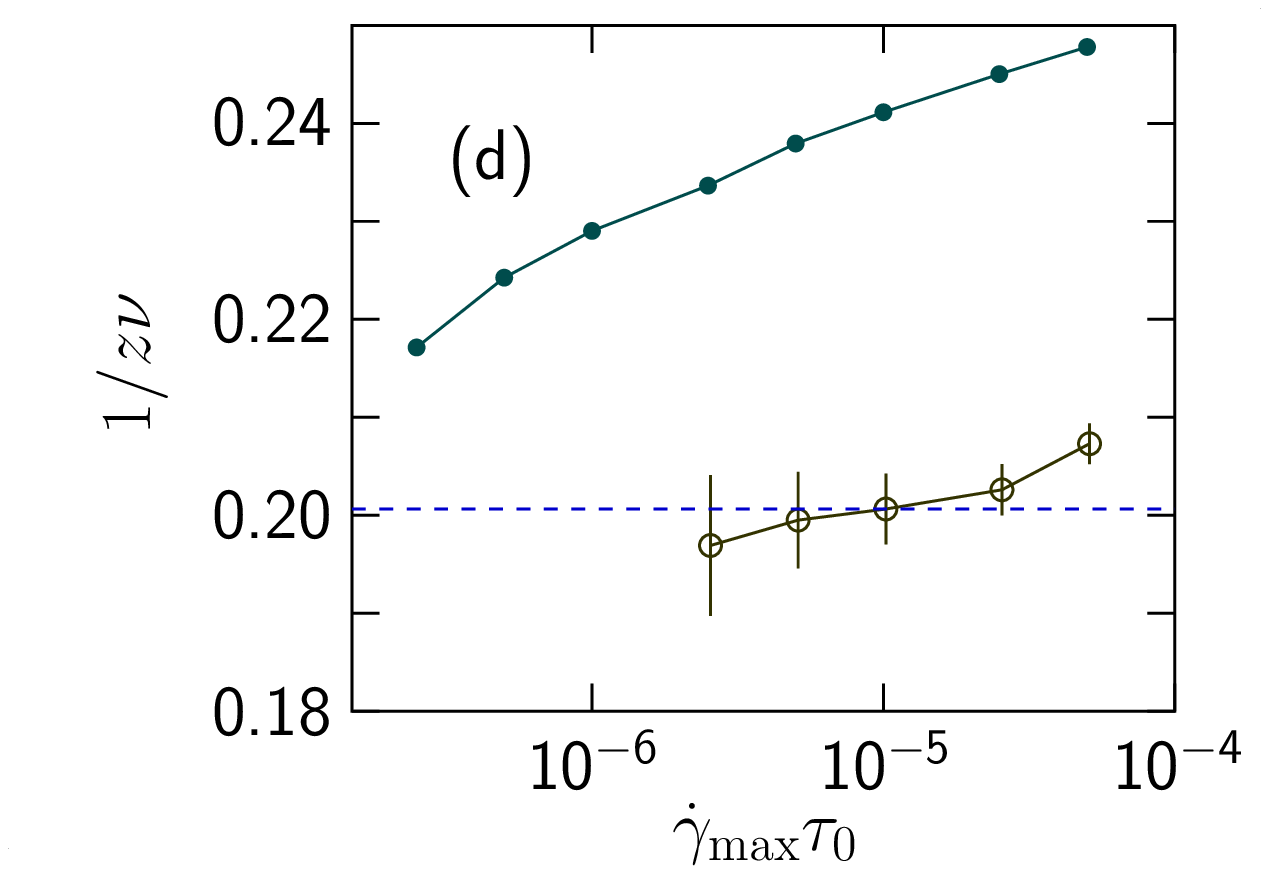}
  \caption{Results from scaling analyses of the pressure. Shown here are (a) the quality
    of the fits in terms of $\chi^2/\mathrm{DOF}$, (b) the exponent $\beta$, (c) $\phi_J$,
    and, (d) $1/z\nu$. All quantities are plotted against $\gdot_\mathrm{max}$ to examine
    whether the fittings are stable against a changing range of shear rates, which would
    be a requirement for a good fit. The solid dots are from fitting to the simple
    \Eq{p-scale-simple} without corrections to scaling whereas open circles are from
    fitting to the full \Eq{p-scale}. The simple fits (solid dots) are clearly
    unsatisfactory as they give bad quality fits and fitting parameters that vary strongly
    with $\gdot_\mathrm{max}$.}
  \label{fig:p-scale}
\end{figure}

We start out by neglecting the corrections-to-scaling term and fitting to the simpler expression,
\begin{equation}
  \label{eq:p-scale-simple}
  p(\delta\phi,\gdot) = \gdot^q f_p\left(\frac{\delta\phi}{\gdot^{1/z\nu}}\right).
\end{equation}
We then adjust $\phi_J$, $q$, and $z\nu$ together with the coefficients of the polynomial
for $f_p$, to get the best possible fit. As we don't know at the outset how big are the shear
rates that can be used in the analysis, we do these fits with different ranges of $\gdot$,
taking $\gdot_\mathrm{min} \leq \gdot\leq\gdot_\mathrm{max}$ with
$\gdot_\mathrm{min}\tau_0=10^{-8}$ and $\gdot_\mathrm{max}\tau_0=2.5\times10^{-7}$ through
$5\times10^{-5}$. The solid dots in \Fig{p-scale} are from these fits. From the quality of
the fits shown as $\chi^2/\mathrm{DOF}$ in panel (a) it is clear that the fits to the
simpler \Eq{p-scale-simple} are good only when the data are restricted to very low shear
rates.

We then include corrections to scaling by fitting to the full expression, \Eq{p-scale},
taking $\omega/z$ and the coefficients of $g_p$ as additional free parameters. As this
expression includes more fitting parameters we need more data in the fits and the analyses
are therefore only done for $\gdot_\mathrm{max}\tau_0\geq2.5\times10^{-6}$. We conclude
that the fit with $\gdot_\mathrm{max}\tau_0=10^{-5}$ gives reliable results by
considering the quality of the fits together with the (weak) dependence on $\gdotmax$.
We thus estimate
$\beta=(1-q)z\nu=3.82\pm0.28$, $\phi_J= 0.6491\pm 0.0003$, $q=0.233\pm 0.016$,
$1/z\nu=0.200\pm 0.011$, $y=1.16\pm0.03$, and $\omega/z=0.30\pm0.06$. The quoted errors
are max/min values (three standard deviations) whereas the error bars in the figures are
$\pm$ one standard deviation. The errors are estimated with Jackknife resampling.  The
value $\beta\approx 3.8$ in 3D is thus clearly different from the 2D value $\beta\approx
2.7$ \cite{Olsson_Teitel:gdot-scale, beta2d}.

\paragraph{Corrections in the analysis of $\tau$}

Due to the curvature of $\tau$ vs $\delta z$ in 2D \cite{Olsson:jam-tau} it was found
important to only make use of data for small $\delta z$ in the determination of
$\beta/u_z$ \cite{Olsson:jam-tau}. It was then found (not shown) that the determined
$\beta/u_z$ increases as the range of $\delta z$ decreases down to $\deltazmax=0.08$, but
then stays stable. Decreasing $\deltazmax$ further only increases the statistical errors.

\begin{figure}
  \includegraphics[bb=31 324 360 584, width=4.2cm]{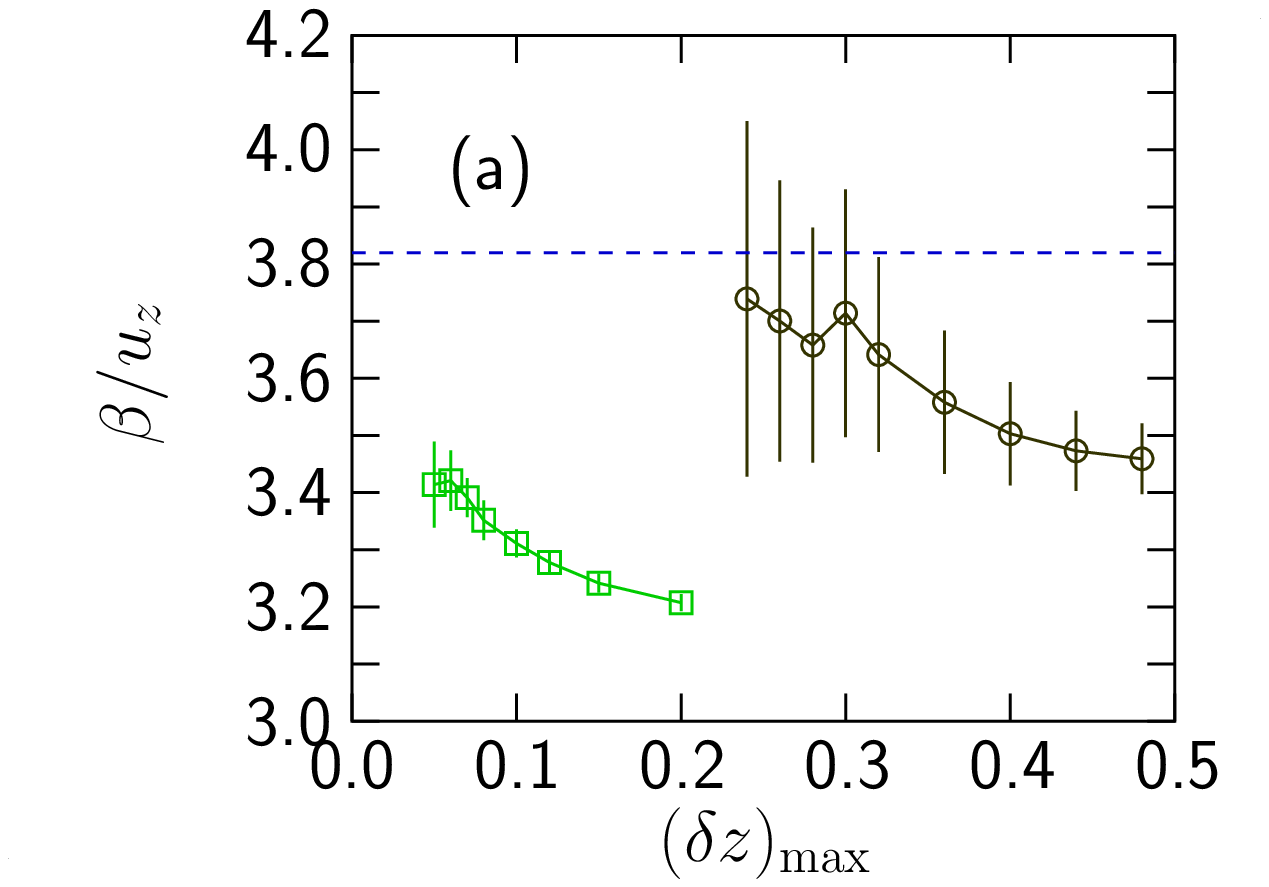}
  \includegraphics[bb=31 324 360 584, width=4.2cm]{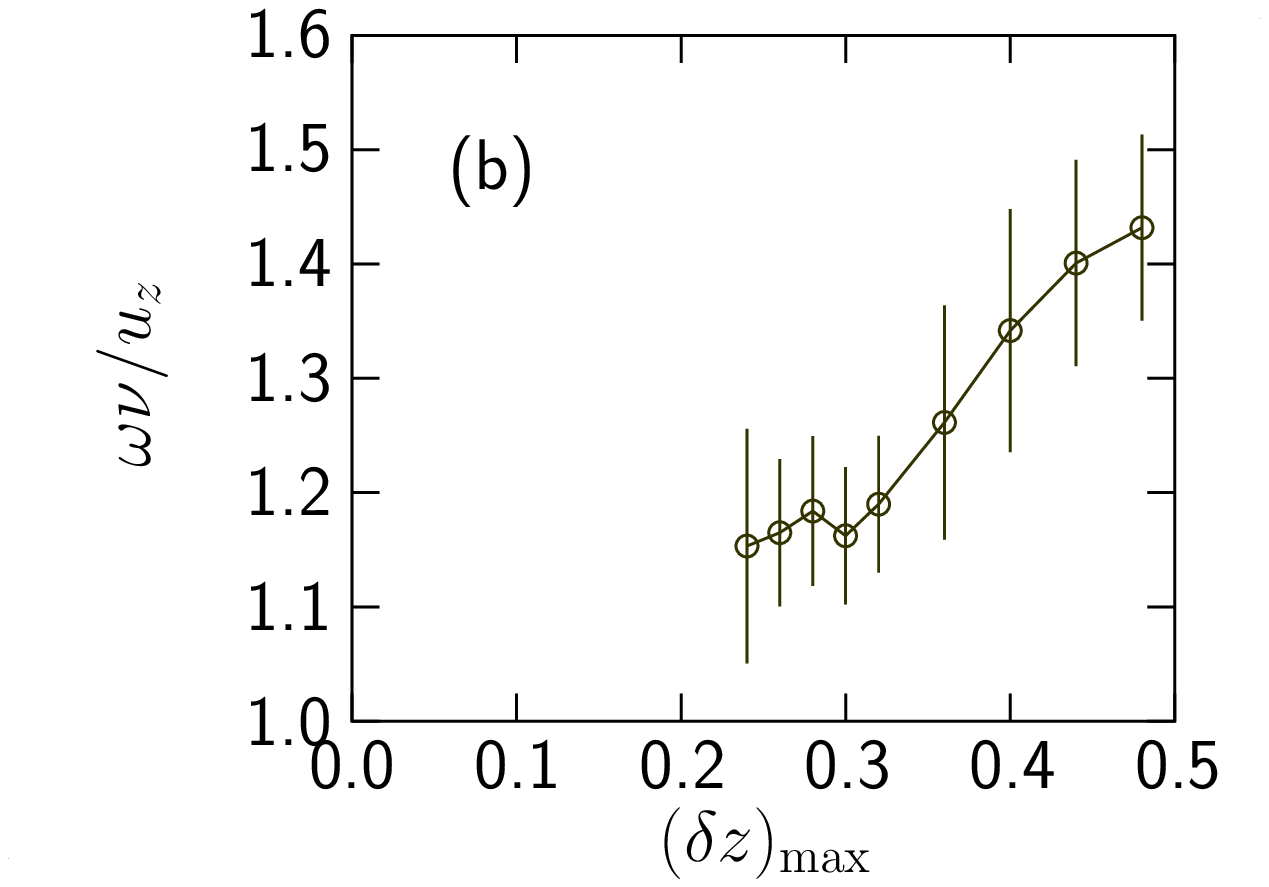}
  \caption{Attempts to refine the determination of $\beta/u_z$ for 3D in
    \Fig{tau-dz}. Panel (a) shows $\beta/u_z$ from fits with different $\deltazmax$. The
    open squares are from linear fits to \Eq{tau} whereas the open circles are from
    fitting $\tau$ to \Eq{tau-dz-corr} that includes corrections to scaling. The dashed
    line is $\beta/u_z=3.82$ with $\beta$ from the scaling analysis shown in
    \Fig{p-scale}, assuming $u_z=1$. Panel (b) shows the correction-to-scaling exponent.}
  \label{fig:tau-dz-corr}
\end{figure}

The analysis of the 3D data in \Fig{tau-dz} was similarly done by fitting to \Eq{tau}
with $\deltazmax=0.08$, and was indeed sufficient for demonstrating that this exponent is
different in 3D compared to 2D. To check for the robustness of this determination, the
lower left part of \Fig{tau-dz-corr}(a) shows $\beta/u_z$ vs $\deltazmax$ for the 3D
data. In contrast to the behavior in 2D, this data does not clearly saturate but rather
gives evidence for a trend to larger $\beta/u_z$ as $\deltazmax$ decreases. The value
$\beta/u_z=3.35$ from \Fig{tau-dz} now only appears as a lower bound.

To try to get a better determination of the 3D exponent we now start from the assumption that
the curvature in $\tau$ vs $\delta z$ is related to corrections to scaling. By
constructing a scaling expression for $p/\gdot$ from \Eq{p-b}, taking $(-\delta\phi)\;
b^{1/\nu}=1$ and $\delta z\sim (-\delta\phi)^{u_z}$, and noting that $\tau(\phi,\gdot\to0)
\sim p(\phi,\gdot\to0)/\gdot$ \cite{Olsson:jam-tau} one arrives at
\begin{equation}
  \label{eq:tau-dz-corr}
  \tau(\delta\phi) = 
  (\delta z)^{-\beta/u_z} \left[ f_0 + (\delta z)^{\omega\nu/u_z} g_0 \right],
\end{equation}
which is \Eq{tau} with a correction term. Similarly to the scaling analysis which was done
for different $\gdotmax$ we fit our data with $\tau$ vs $\delta z$ to \Eq{tau-dz-corr} for
$\delta z \leq\deltazmax$. As shown in \Fig{tau-dz-corr}(a), decreasing the range of data
from $\deltazmax=0.48$ through 0.24 gives evidence for trends in both $\beta/u_z$ and
$\omega\nu/u_z$ which appear to saturate at $\deltazmax=0.30$. We therefore read off
$\beta/u_z=3.7\pm0.7$. We also note that this value appears as a reasonable candidate to an
extrapolation of the open squares in \Fig{tau-dz-corr}(a) to $\deltazmax=0$.  Since
we find numerically that $\beta$ from $p$ is equal to $\beta/u_z$ from $\tau$, we conclude
$u_z\approx1$ in 3D, in agreement with the above-mentioned $u_z\approx1$ in 2D.

\paragraph{Comparison with the literature}

Evidence for differing exponents in two and three dimensions has actually for some time
been available in the literature. The first determinations of $\beta/u_z$ (there denoted
by $1/\delta$) in \Ref{Lerner-PNAS:2012}, gave $\beta/u_z=1/0.38 = 2.63$ in 2D and
$\beta/u_z=1/0.34=2.94$ in 3D. Those authors, however, did not consider this a significant
difference. The main source of uncertainty in these analyses is whether the data are
sufficiently close to criticality to give the true critical behavior. In a later paper by
the same group \cite{DeGiuli:2015}, simulations closer to criticality---i.e.\ at smaller
$\delta z$---gave $\beta/u_z=1/0.3\approx3.3$ in 3D, but as that paper was focused on
comparisons with theory they didn't comment on possible differences between two and three
dimensions. Their values do however agree nicely with our analyses in \Fig{tau-dz}.

We also note from \Fig{p-scale}(b) that a simple scaling analysis of $p(\phi,\gdot)$,
based on \Eq{p-scale-simple}, without corrections to scaling, gives $\beta\approx 2.9$,
close to the 2D value, provided that one includes in the fit larger values of the strain
rate $\gdot$. The bad quality of the fit in \Fig{p-scale}(a) makes clear that these low
values cannot be correct but it shows that analyses may give the erroneous conclusion that
the critical behavior in 3D and 2D are the same, seemingly confirming the prevailing
paradigm.  
Another example of a low value in the literature is $\beta\approx 1/0.391= 2.56$
\cite{Kawasaki_Berthier:2015}. This, again, appears to be an effect of using data too far
from criticality, as the fits, according to their Fig.~4(a), include points for densities
down to, or below, $\phi\approx\phi_J-0.05$. To compare, the data used in the scaling
analyses in the present work are restricted to $|\phi-\phi_J|\leq 0.017$.

\paragraph{Discussion}

Recent attempts by the group of Wyart to determine the exponents analytically, in terms of
the exponent $\theta_e$, rely on examining the properties of the opening and closing of
contacts \cite{DeGiuli:2015, During_Lerner_Wyart:2016}. The exponent $\theta_e$
characterizes the distribution of weak forces in packings from isotropic jamming and has
been found to be $\theta_e\approx0.42311$ by analytic calculations in infinite dimensions
\cite{Charbonneau:NatCommun:2014, Charbonneau:Jstat:2014}. It is also found to be the same
in 2D and 3D \cite{Lerner-SoftMatter:2013, DeGiuli:PNAS:2014, Charbonneau:2015:prl} and is
believed \cite{DeGiuli:2015} to be the same also in the shear-driven case.

The result of the present Letter, that critical exponents for shear-driven jamming are
different for 3D compared to 2D, is however in conflict with a picture where the exponents
only depend on the dimension-independent exponent $\theta_e$. One possible reason for this
difference could be that their relations \cite{DeGiuli:2015, During_Lerner_Wyart:2016}
describe the typical particle motion whereas the dissipation (and the viscosity)
is instead dominated by a small fraction of particles with the highest velocity---a
fraction which decreases as jamming is approached \cite{Olsson:jam-vhist}. The
investigation into this issue appears as an important direction for future research.

\paragraph{Conclusion}

From shear-driven simulations of elastic particles in three dimensions together with
previous results for two dimensions, we determine the critical exponents of the
shear-driven jamming of frictionless athermal particles and conclude---in variance with
the prevailing picture---that the 3D and 2D transitions do not belong to the same
universality class.

\begin{acknowledgments}
  I thank S. Teitel for suggestions, discussions, and a critical reading of the
  manuscript.  Simulations were performed on resources provided by the Swedish National
  Infrastructure for Computing (SNIC) at HPC2N.
\end{acknowledgments}

\bibliography{j,this}

\begin{thebibliography}{32}%
\makeatletter
\providecommand \@ifxundefined [1]{%
 \@ifx{#1\undefined}
}%
\providecommand \@ifnum [1]{%
 \ifnum #1\expandafter \@firstoftwo
 \else \expandafter \@secondoftwo
 \fi
}%
\providecommand \@ifx [1]{%
 \ifx #1\expandafter \@firstoftwo
 \else \expandafter \@secondoftwo
 \fi
}%
\providecommand \natexlab [1]{#1}%
\providecommand \enquote  [1]{``#1''}%
\providecommand \bibnamefont  [1]{#1}%
\providecommand \bibfnamefont [1]{#1}%
\providecommand \citenamefont [1]{#1}%
\providecommand \href@noop [0]{\@secondoftwo}%
\providecommand \href [0]{\begingroup \@sanitize@url \@href}%
\providecommand \@href[1]{\@@startlink{#1}\@@href}%
\providecommand \@@href[1]{\endgroup#1\@@endlink}%
\providecommand \@sanitize@url [0]{\catcode `\\12\catcode `\$12\catcode
  `\&12\catcode `\#12\catcode `\^12\catcode `\_12\catcode `\%12\relax}%
\providecommand \@@startlink[1]{}%
\providecommand \@@endlink[0]{}%
\providecommand \url  [0]{\begingroup\@sanitize@url \@url }%
\providecommand \@url [1]{\endgroup\@href {#1}{\urlprefix }}%
\providecommand \urlprefix  [0]{URL }%
\providecommand \Eprint [0]{\href }%
\providecommand \doibase [0]{http://dx.doi.org/}%
\providecommand \selectlanguage [0]{\@gobble}%
\providecommand \bibinfo  [0]{\@secondoftwo}%
\providecommand \bibfield  [0]{\@secondoftwo}%
\providecommand \translation [1]{[#1]}%
\providecommand \BibitemOpen [0]{}%
\providecommand \bibitemStop [0]{}%
\providecommand \bibitemNoStop [0]{.\EOS\space}%
\providecommand \EOS [0]{\spacefactor3000\relax}%
\providecommand \BibitemShut  [1]{\csname bibitem#1\endcsname}%
\let\auto@bib@innerbib\@empty
\bibitem [{\citenamefont {Donev}\ \emph {et~al.}(2005)\citenamefont {Donev},
  \citenamefont {Torquato},\ and\ \citenamefont
  {Stillinger}}]{Donev_TS:packing}%
  \BibitemOpen
  \bibfield  {author} {\bibinfo {author} {\bibfnamefont {A.}~\bibnamefont
  {Donev}}, \bibinfo {author} {\bibfnamefont {S.}~\bibnamefont {Torquato}}, \
  and\ \bibinfo {author} {\bibfnamefont {F.~H.}\ \bibnamefont {Stillinger}},\
  }\href {\doibase 10.1103/PhysRevE.71.011105} {\bibfield  {journal} {\bibinfo
  {journal} {Phys. Rev. E}\ }\textbf {\bibinfo {volume} {71}},\ \bibinfo
  {pages} {011105} (\bibinfo {year} {2005})}\BibitemShut {NoStop}%
\bibitem [{\citenamefont {Berthier}\ and\ \citenamefont
  {Witten}(2009)}]{Berthier_Witten:PRE2009}%
  \BibitemOpen
  \bibfield  {author} {\bibinfo {author} {\bibfnamefont {L.}~\bibnamefont
  {Berthier}}\ and\ \bibinfo {author} {\bibfnamefont {T.~A.}\ \bibnamefont
  {Witten}},\ }\href {\doibase 10.1103/PhysRevE.80.021502} {\bibfield
  {journal} {\bibinfo  {journal} {Phys. Rev. E}\ }\textbf {\bibinfo {volume}
  {80}},\ \bibinfo {pages} {021502} (\bibinfo {year} {2009})}\BibitemShut
  {NoStop}%
\bibitem [{\citenamefont {V{\aa}gberg}\ \emph {et~al.}(2011)\citenamefont
  {V{\aa}gberg}, \citenamefont {Olsson},\ and\ \citenamefont
  {Teitel}}]{Vagberg_OT:protocol}%
  \BibitemOpen
  \bibfield  {author} {\bibinfo {author} {\bibfnamefont {D.}~\bibnamefont
  {V{\aa}gberg}}, \bibinfo {author} {\bibfnamefont {P.}~\bibnamefont {Olsson}},
  \ and\ \bibinfo {author} {\bibfnamefont {S.}~\bibnamefont {Teitel}},\ }\href
  {\doibase 10.1103/PhysRevE.83.031307} {\bibfield  {journal} {\bibinfo
  {journal} {Phys.\ Rev.\ E}\ }\textbf {\bibinfo {volume} {83}},\ \bibinfo
  {pages} {031307} (\bibinfo {year} {2011})}\BibitemShut {NoStop}%
\bibitem [{\citenamefont {Ozawa}\ \emph {et~al.}(2017)\citenamefont {Ozawa},
  \citenamefont {Berthier},\ and\ \citenamefont {Coslovich}}]{Ozawa:2017}%
  \BibitemOpen
  \bibfield  {author} {\bibinfo {author} {\bibfnamefont {M.}~\bibnamefont
  {Ozawa}}, \bibinfo {author} {\bibfnamefont {L.}~\bibnamefont {Berthier}}, \
  and\ \bibinfo {author} {\bibfnamefont {D.}~\bibnamefont {Coslovich}},\
  }\href@noop {} {\bibfield  {journal} {\bibinfo  {journal} {SciPost Phys.}\
  }\textbf {\bibinfo {volume} {3}},\ \bibinfo {pages} {027} (\bibinfo {year}
  {2017})}\BibitemShut {NoStop}%
\bibitem [{\citenamefont {O'Hern}\ \emph {et~al.}(2002)\citenamefont {O'Hern},
  \citenamefont {Langer}, \citenamefont {Liu},\ and\ \citenamefont
  {Nagel}}]{OHern_Langer_Liu_Nagel:2002}%
  \BibitemOpen
  \bibfield  {author} {\bibinfo {author} {\bibfnamefont {C.~S.}\ \bibnamefont
  {O'Hern}}, \bibinfo {author} {\bibfnamefont {S.~A.}\ \bibnamefont {Langer}},
  \bibinfo {author} {\bibfnamefont {A.~J.}\ \bibnamefont {Liu}}, \ and\
  \bibinfo {author} {\bibfnamefont {S.~R.}\ \bibnamefont {Nagel}},\ }\href@noop
  {} {\bibfield  {journal} {\bibinfo  {journal} {Phys. Rev. Lett.}\ }\textbf
  {\bibinfo {volume} {88}},\ \bibinfo {pages} {075507} (\bibinfo {year}
  {2002})}\BibitemShut {NoStop}%
\bibitem [{\citenamefont {O'Hern}\ \emph {et~al.}(2003)\citenamefont {O'Hern},
  \citenamefont {Silbert}, \citenamefont {Liu},\ and\ \citenamefont
  {Nagel}}]{OHern_Silbert_Liu_Nagel:2003}%
  \BibitemOpen
  \bibfield  {author} {\bibinfo {author} {\bibfnamefont {C.~S.}\ \bibnamefont
  {O'Hern}}, \bibinfo {author} {\bibfnamefont {L.~E.}\ \bibnamefont {Silbert}},
  \bibinfo {author} {\bibfnamefont {A.~J.}\ \bibnamefont {Liu}}, \ and\
  \bibinfo {author} {\bibfnamefont {S.~R.}\ \bibnamefont {Nagel}},\ }\href
  {\doibase 10.1103/PhysRevE.68.011306} {\bibfield  {journal} {\bibinfo
  {journal} {Phys. Rev. E}\ }\textbf {\bibinfo {volume} {68}},\ \bibinfo
  {pages} {011306} (\bibinfo {year} {2003})}\BibitemShut {NoStop}%
\bibitem [{\citenamefont {Chaudhuri}\ \emph {et~al.}(2010)\citenamefont
  {Chaudhuri}, \citenamefont {Berthier},\ and\ \citenamefont
  {Sastry}}]{Chaudhuri_Berthier_Sastry}%
  \BibitemOpen
  \bibfield  {author} {\bibinfo {author} {\bibfnamefont {P.}~\bibnamefont
  {Chaudhuri}}, \bibinfo {author} {\bibfnamefont {L.}~\bibnamefont {Berthier}},
  \ and\ \bibinfo {author} {\bibfnamefont {S.}~\bibnamefont {Sastry}},\ }\href
  {\doibase 10.1103/PhysRevLett.104.165701} {\bibfield  {journal} {\bibinfo
  {journal} {Phys.\ Rev.\ Lett.}\ }\textbf {\bibinfo {volume} {104}},\ \bibinfo
  {pages} {165701} (\bibinfo {year} {2010})}\BibitemShut {NoStop}%
\bibitem [{\citenamefont {Lerner}\ \emph {et~al.}(2013)\citenamefont {Lerner},
  \citenamefont {During},\ and\ \citenamefont
  {Wyart}}]{Lerner-SoftMatter:2013}%
  \BibitemOpen
  \bibfield  {author} {\bibinfo {author} {\bibfnamefont {E.}~\bibnamefont
  {Lerner}}, \bibinfo {author} {\bibfnamefont {G.}~\bibnamefont {During}}, \
  and\ \bibinfo {author} {\bibfnamefont {M.}~\bibnamefont {Wyart}},\ }\href
  {\doibase 10.1039/C3SM50515D} {\bibfield  {journal} {\bibinfo  {journal}
  {Soft Matter}\ }\textbf {\bibinfo {volume} {9}},\ \bibinfo {pages} {8252}
  (\bibinfo {year} {2013})}\BibitemShut {NoStop}%
\bibitem [{\citenamefont {DeGiuli}\ \emph {et~al.}(2014)\citenamefont
  {DeGiuli}, \citenamefont {Lerner}, \citenamefont {Brito},\ and\ \citenamefont
  {Wyart}}]{DeGiuli:PNAS:2014}%
  \BibitemOpen
  \bibfield  {author} {\bibinfo {author} {\bibfnamefont {E.}~\bibnamefont
  {DeGiuli}}, \bibinfo {author} {\bibfnamefont {E.}~\bibnamefont {Lerner}},
  \bibinfo {author} {\bibfnamefont {C.}~\bibnamefont {Brito}}, \ and\ \bibinfo
  {author} {\bibfnamefont {M.}~\bibnamefont {Wyart}},\ }\href {\doibase
  10.1073/pnas.1415298111} {\bibfield  {journal} {\bibinfo  {journal}
  {Proceedings of the National Academy of Sciences}\ }\textbf {\bibinfo
  {volume} {111}},\ \bibinfo {pages} {17054} (\bibinfo {year}
  {2014})}\BibitemShut {NoStop}%
\bibitem [{\citenamefont {Charbonneau}\ \emph {et~al.}(2015)\citenamefont
  {Charbonneau}, \citenamefont {Corwin}, \citenamefont {Parisi},\ and\
  \citenamefont {Zamponi}}]{Charbonneau:2015:prl}%
  \BibitemOpen
  \bibfield  {author} {\bibinfo {author} {\bibfnamefont {P.}~\bibnamefont
  {Charbonneau}}, \bibinfo {author} {\bibfnamefont {E.~I.}\ \bibnamefont
  {Corwin}}, \bibinfo {author} {\bibfnamefont {G.}~\bibnamefont {Parisi}}, \
  and\ \bibinfo {author} {\bibfnamefont {F.}~\bibnamefont {Zamponi}},\ }\href
  {\doibase 10.1103/PhysRevLett.114.125504} {\bibfield  {journal} {\bibinfo
  {journal} {Phys. Rev. Lett.}\ }\textbf {\bibinfo {volume} {114}},\ \bibinfo
  {pages} {125504} (\bibinfo {year} {2015})}\BibitemShut {NoStop}%
\bibitem [{\citenamefont {Charbonneau}\ \emph
  {et~al.}(2014{\natexlab{a}})\citenamefont {Charbonneau}, \citenamefont
  {Kurchan}, \citenamefont {Parisi}, \citenamefont {Urbani},\ and\
  \citenamefont {Zamponi}}]{Charbonneau:NatCommun:2014}%
  \BibitemOpen
  \bibfield  {author} {\bibinfo {author} {\bibfnamefont {P.}~\bibnamefont
  {Charbonneau}}, \bibinfo {author} {\bibfnamefont {J.}~\bibnamefont
  {Kurchan}}, \bibinfo {author} {\bibfnamefont {G.}~\bibnamefont {Parisi}},
  \bibinfo {author} {\bibfnamefont {P.}~\bibnamefont {Urbani}}, \ and\ \bibinfo
  {author} {\bibfnamefont {F.}~\bibnamefont {Zamponi}},\ }\href@noop {}
  {\bibfield  {journal} {\bibinfo  {journal} {Nature Communications}\ }\textbf
  {\bibinfo {volume} {5}},\ \bibinfo {pages} {3725} (\bibinfo {year}
  {2014}{\natexlab{a}})}\BibitemShut {NoStop}%
\bibitem [{\citenamefont {Charbonneau}\ \emph
  {et~al.}(2014{\natexlab{b}})\citenamefont {Charbonneau}, \citenamefont
  {Kurchan}, \citenamefont {Parisi}, \citenamefont {Urbani},\ and\
  \citenamefont {Zamponi}}]{Charbonneau:Jstat:2014}%
  \BibitemOpen
  \bibfield  {author} {\bibinfo {author} {\bibfnamefont {P.}~\bibnamefont
  {Charbonneau}}, \bibinfo {author} {\bibfnamefont {J.}~\bibnamefont
  {Kurchan}}, \bibinfo {author} {\bibfnamefont {G.}~\bibnamefont {Parisi}},
  \bibinfo {author} {\bibfnamefont {P.}~\bibnamefont {Urbani}}, \ and\ \bibinfo
  {author} {\bibfnamefont {F.}~\bibnamefont {Zamponi}},\ }\href
  {http://stacks.iop.org/1742-5468/2014/i=10/a=P10009} {\bibfield  {journal}
  {\bibinfo  {journal} {Journal of Statistical Mechanics: Theory and
  Experiment}\ }\textbf {\bibinfo {volume} {2014}},\ \bibinfo {pages} {P10009}
  (\bibinfo {year} {2014}{\natexlab{b}})}\BibitemShut {NoStop}%
\bibitem [{\citenamefont {Wyart}\ \emph {et~al.}(2005)\citenamefont {Wyart},
  \citenamefont {Silbert}, \citenamefont {Nagel},\ and\ \citenamefont
  {Witten}}]{Wyart:2005}%
  \BibitemOpen
  \bibfield  {author} {\bibinfo {author} {\bibfnamefont {M.}~\bibnamefont
  {Wyart}}, \bibinfo {author} {\bibfnamefont {L.~E.}\ \bibnamefont {Silbert}},
  \bibinfo {author} {\bibfnamefont {S.~R.}\ \bibnamefont {Nagel}}, \ and\
  \bibinfo {author} {\bibfnamefont {T.~A.}\ \bibnamefont {Witten}},\ }\href
  {\doibase 10.1103/PhysRevE.72.051306} {\bibfield  {journal} {\bibinfo
  {journal} {Phys. Rev. E}\ }\textbf {\bibinfo {volume} {72}},\ \bibinfo
  {pages} {051306} (\bibinfo {year} {2005})}\BibitemShut {NoStop}%
\bibitem [{\citenamefont {Goodrich}\ \emph {et~al.}(2012)\citenamefont
  {Goodrich}, \citenamefont {Liu},\ and\ \citenamefont
  {Nagel}}]{Goodrich:2012}%
  \BibitemOpen
  \bibfield  {author} {\bibinfo {author} {\bibfnamefont {C.~P.}\ \bibnamefont
  {Goodrich}}, \bibinfo {author} {\bibfnamefont {A.~J.}\ \bibnamefont {Liu}}, \
  and\ \bibinfo {author} {\bibfnamefont {S.~R.}\ \bibnamefont {Nagel}},\ }\href
  {\doibase 10.1103/PhysRevLett.109.095704} {\bibfield  {journal} {\bibinfo
  {journal} {Phys. Rev. Lett.}\ }\textbf {\bibinfo {volume} {109}},\ \bibinfo
  {pages} {095704} (\bibinfo {year} {2012})}\BibitemShut {NoStop}%
\bibitem [{\citenamefont {DeGiuli}\ \emph {et~al.}(2015)\citenamefont
  {DeGiuli}, \citenamefont {D\"uring}, \citenamefont {Lerner},\ and\
  \citenamefont {Wyart}}]{DeGiuli:2015}%
  \BibitemOpen
  \bibfield  {author} {\bibinfo {author} {\bibfnamefont {E.}~\bibnamefont
  {DeGiuli}}, \bibinfo {author} {\bibfnamefont {G.}~\bibnamefont {D\"uring}},
  \bibinfo {author} {\bibfnamefont {E.}~\bibnamefont {Lerner}}, \ and\ \bibinfo
  {author} {\bibfnamefont {M.}~\bibnamefont {Wyart}},\ }\href {\doibase
  10.1103/PhysRevE.91.062206} {\bibfield  {journal} {\bibinfo  {journal} {Phys.
  Rev. E}\ }\textbf {\bibinfo {volume} {91}},\ \bibinfo {pages} {062206}
  (\bibinfo {year} {2015})}\BibitemShut {NoStop}%
\bibitem [{\citenamefont {D\"uring}\ \emph {et~al.}(2016)\citenamefont
  {D\"uring}, \citenamefont {Lerner},\ and\ \citenamefont
  {Wyart}}]{During_Lerner_Wyart:2016}%
  \BibitemOpen
  \bibfield  {author} {\bibinfo {author} {\bibfnamefont {G.}~\bibnamefont
  {D\"uring}}, \bibinfo {author} {\bibfnamefont {E.}~\bibnamefont {Lerner}}, \
  and\ \bibinfo {author} {\bibfnamefont {M.}~\bibnamefont {Wyart}},\ }\href
  {\doibase 10.1103/PhysRevE.94.022601} {\bibfield  {journal} {\bibinfo
  {journal} {Phys. Rev. E}\ }\textbf {\bibinfo {volume} {94}},\ \bibinfo
  {pages} {022601} (\bibinfo {year} {2016})}\BibitemShut {NoStop}%
\bibitem [{\citenamefont {Lerner}\ \emph {et~al.}(2012)\citenamefont {Lerner},
  \citenamefont {D\"{u}ring},\ and\ \citenamefont {Wyart}}]{Lerner-PNAS:2012}%
  \BibitemOpen
  \bibfield  {author} {\bibinfo {author} {\bibfnamefont {E.}~\bibnamefont
  {Lerner}}, \bibinfo {author} {\bibfnamefont {G.}~\bibnamefont {D\"{u}ring}},
  \ and\ \bibinfo {author} {\bibfnamefont {M.}~\bibnamefont {Wyart}},\
  }\href@noop {} {\bibfield  {journal} {\bibinfo  {journal} {PNAS}\ }\textbf
  {\bibinfo {volume} {109}},\ \bibinfo {pages} {4798} (\bibinfo {year}
  {2012})}\BibitemShut {NoStop}%
\bibitem [{\citenamefont {Olsson}\ and\ \citenamefont
  {Teitel}(2013)}]{Olsson_Teitel:jam-T}%
  \BibitemOpen
  \bibfield  {author} {\bibinfo {author} {\bibfnamefont {P.}~\bibnamefont
  {Olsson}}\ and\ \bibinfo {author} {\bibfnamefont {S.}~\bibnamefont
  {Teitel}},\ }\href {\doibase 10.1103/PhysRevE.88.010301} {\bibfield
  {journal} {\bibinfo  {journal} {Phys. Rev. E}\ }\textbf {\bibinfo {volume}
  {88}},\ \bibinfo {pages} {010301} (\bibinfo {year} {2013})}\BibitemShut
  {NoStop}%
\bibitem [{\citenamefont {Olsson}\ and\ \citenamefont
  {Teitel}(2007)}]{Olsson_Teitel:jamming}%
  \BibitemOpen
  \bibfield  {author} {\bibinfo {author} {\bibfnamefont {P.}~\bibnamefont
  {Olsson}}\ and\ \bibinfo {author} {\bibfnamefont {S.}~\bibnamefont
  {Teitel}},\ }\href {\doibase 10.1103/PhysRevLett.99.178001} {\bibfield
  {journal} {\bibinfo  {journal} {Phys. Rev. Lett.}\ }\textbf {\bibinfo
  {volume} {99}},\ \bibinfo {pages} {178001} (\bibinfo {year}
  {2007})}\BibitemShut {NoStop}%
\bibitem [{\citenamefont {Hatano}(2008)}]{Hatano:2008}%
  \BibitemOpen
  \bibfield  {author} {\bibinfo {author} {\bibfnamefont {T.}~\bibnamefont
  {Hatano}},\ }\href@noop {} {\bibfield  {journal} {\bibinfo  {journal} {J.
  Phys. Soc. Jpn.}\ }\textbf {\bibinfo {volume} {77}},\ \bibinfo {pages}
  {123002} (\bibinfo {year} {2008})}\BibitemShut {NoStop}%
\bibitem [{\citenamefont {Heussinger}\ and\ \citenamefont
  {Barrat}(2009)}]{Heussinger_Barrat:2009}%
  \BibitemOpen
  \bibfield  {author} {\bibinfo {author} {\bibfnamefont {C.}~\bibnamefont
  {Heussinger}}\ and\ \bibinfo {author} {\bibfnamefont {J.-L.}\ \bibnamefont
  {Barrat}},\ }\href {\doibase 10.1103/PhysRevLett.102.218303} {\bibfield
  {journal} {\bibinfo  {journal} {Phys. Rev. Lett.}\ }\textbf {\bibinfo
  {volume} {102}},\ \bibinfo {pages} {218303} (\bibinfo {year}
  {2009})}\BibitemShut {NoStop}%
\bibitem [{\citenamefont {Otsuki}\ and\ \citenamefont
  {Hayakawa}(2009)}]{Otsuki_Hayakawa:2009b}%
  \BibitemOpen
  \bibfield  {author} {\bibinfo {author} {\bibfnamefont {M.}~\bibnamefont
  {Otsuki}}\ and\ \bibinfo {author} {\bibfnamefont {H.}~\bibnamefont
  {Hayakawa}},\ }\href {\doibase 10.1103/PhysRevE.80.011308} {\bibfield
  {journal} {\bibinfo  {journal} {Phys. Rev. E}\ }\textbf {\bibinfo {volume}
  {80}},\ \bibinfo {pages} {011308} (\bibinfo {year} {2009})}\BibitemShut
  {NoStop}%
\bibitem [{\citenamefont {Hatano}(2010)}]{Hatano:2010}%
  \BibitemOpen
  \bibfield  {author} {\bibinfo {author} {\bibfnamefont {T.}~\bibnamefont
  {Hatano}},\ }\href {\doibase 10.1143/PTPS.184.143} {\bibfield  {journal}
  {\bibinfo  {journal} {Prog.\ Theor.\ Phys.\ Suppl.}\ }\textbf {\bibinfo
  {volume} {184}},\ \bibinfo {pages} {143} (\bibinfo {year}
  {2010})}\BibitemShut {NoStop}%
\bibitem [{\citenamefont {Tighe}\ \emph {et~al.}(2010)\citenamefont {Tighe},
  \citenamefont {Woldhuis}, \citenamefont {Remmers}, \citenamefont {van
  Saarloos},\ and\ \citenamefont {van Hecke}}]{Tighe_WRvSvH}%
  \BibitemOpen
  \bibfield  {author} {\bibinfo {author} {\bibfnamefont {B.~P.}\ \bibnamefont
  {Tighe}}, \bibinfo {author} {\bibfnamefont {E.}~\bibnamefont {Woldhuis}},
  \bibinfo {author} {\bibfnamefont {J.~J.~C.}\ \bibnamefont {Remmers}},
  \bibinfo {author} {\bibfnamefont {W.}~\bibnamefont {van Saarloos}}, \ and\
  \bibinfo {author} {\bibfnamefont {M.}~\bibnamefont {van Hecke}},\ }\href
  {\doibase 10.1103/PhysRevLett.105.088303} {\bibfield  {journal} {\bibinfo
  {journal} {Phys. Rev. Lett.}\ }\textbf {\bibinfo {volume} {105}},\ \bibinfo
  {pages} {088303} (\bibinfo {year} {2010})}\BibitemShut {NoStop}%
\bibitem [{\citenamefont {Olsson}\ and\ \citenamefont
  {Teitel}(2011)}]{Olsson_Teitel:gdot-scale}%
  \BibitemOpen
  \bibfield  {author} {\bibinfo {author} {\bibfnamefont {P.}~\bibnamefont
  {Olsson}}\ and\ \bibinfo {author} {\bibfnamefont {S.}~\bibnamefont
  {Teitel}},\ }\href {\doibase 10.1103/PhysRevE.83.030302} {\bibfield
  {journal} {\bibinfo  {journal} {Phys.\ Rev.\ E}\ }\textbf {\bibinfo {volume}
  {83}},\ \bibinfo {pages} {030302(R)} (\bibinfo {year} {2011})}\BibitemShut
  {NoStop}%
\bibitem [{\citenamefont {Kawasaki}\ \emph {et~al.}(2015)\citenamefont
  {Kawasaki}, \citenamefont {Coslovich}, \citenamefont {Ikeda},\ and\
  \citenamefont {Berthier}}]{Kawasaki_Berthier:2015}%
  \BibitemOpen
  \bibfield  {author} {\bibinfo {author} {\bibfnamefont {T.}~\bibnamefont
  {Kawasaki}}, \bibinfo {author} {\bibfnamefont {D.}~\bibnamefont {Coslovich}},
  \bibinfo {author} {\bibfnamefont {A.}~\bibnamefont {Ikeda}}, \ and\ \bibinfo
  {author} {\bibfnamefont {L.}~\bibnamefont {Berthier}},\ }\href {\doibase
  10.1103/PhysRevE.91.012203} {\bibfield  {journal} {\bibinfo  {journal} {Phys.
  Rev. E}\ }\textbf {\bibinfo {volume} {91}},\ \bibinfo {pages} {012203}
  (\bibinfo {year} {2015})}\BibitemShut {NoStop}%
\bibitem [{bet()}]{beta2d}%
  \BibitemOpen
  \href@noop {} {}\bibinfo {note} {The exponents in
  Ref.~\cite{Olsson_Teitel:gdot-scale} give $\beta=(1-q)z\nu\approx 2.64$. A
  more recent determination with the same method but more precise data gives
  $\beta=2.70\pm0.15$.}\BibitemShut {Stop}%
\bibitem [{\citenamefont {Olsson}(2015)}]{Olsson:jam-tau}%
  \BibitemOpen
  \bibfield  {author} {\bibinfo {author} {\bibfnamefont {P.}~\bibnamefont
  {Olsson}},\ }\href {\doibase 10.1103/PhysRevE.91.062209} {\bibfield
  {journal} {\bibinfo  {journal} {Phys. Rev. E}\ }\textbf {\bibinfo {volume}
  {91}},\ \bibinfo {pages} {062209} (\bibinfo {year} {2015})},\ \bibinfo {note}
  {the error estimate given here is obtained from a jackknife analysis of the
  original data}\BibitemShut {NoStop}%
\bibitem [{\citenamefont {Brambilla}\ \emph {et~al.}(2009)\citenamefont
  {Brambilla}, \citenamefont {El~Masri}, \citenamefont {Pierno}, \citenamefont
  {Berthier}, \citenamefont {Cipelletti}, \citenamefont {Petekidis},\ and\
  \citenamefont {Schofield}}]{Brambilla:2009}%
  \BibitemOpen
  \bibfield  {author} {\bibinfo {author} {\bibfnamefont {G.}~\bibnamefont
  {Brambilla}}, \bibinfo {author} {\bibfnamefont {D.}~\bibnamefont {El~Masri}},
  \bibinfo {author} {\bibfnamefont {M.}~\bibnamefont {Pierno}}, \bibinfo
  {author} {\bibfnamefont {L.}~\bibnamefont {Berthier}}, \bibinfo {author}
  {\bibfnamefont {L.}~\bibnamefont {Cipelletti}}, \bibinfo {author}
  {\bibfnamefont {G.}~\bibnamefont {Petekidis}}, \ and\ \bibinfo {author}
  {\bibfnamefont {A.~B.}\ \bibnamefont {Schofield}},\ }\href {\doibase
  10.1103/PhysRevLett.102.085703} {\bibfield  {journal} {\bibinfo  {journal}
  {Phys. Rev. Lett.}\ }\textbf {\bibinfo {volume} {102}},\ \bibinfo {pages}
  {085703} (\bibinfo {year} {2009})}\BibitemShut {NoStop}%
\bibitem [{\citenamefont {Evans}\ and\ \citenamefont
  {Morriss}(1990)}]{Evans_Morriss}%
  \BibitemOpen
  \bibfield  {author} {\bibinfo {author} {\bibfnamefont {D.~J.}\ \bibnamefont
  {Evans}}\ and\ \bibinfo {author} {\bibfnamefont {G.~P.}\ \bibnamefont
  {Morriss}},\ }\href@noop {} {\emph {\bibinfo {title} {Statistical Mechanics
  of Nonequilibrium Liquids}}}\ (\bibinfo  {publisher} {Academic Press},\
  \bibinfo {address} {London},\ \bibinfo {year} {1990})\BibitemShut {NoStop}%
\bibitem [{\citenamefont {V{\aa}gberg}\ \emph {et~al.}(2014)\citenamefont
  {V{\aa}gberg}, \citenamefont {Olsson},\ and\ \citenamefont
  {Teitel}}]{Vagberg_Olsson_Teitel:BagnNewt}%
  \BibitemOpen
  \bibfield  {author} {\bibinfo {author} {\bibfnamefont {D.}~\bibnamefont
  {V{\aa}gberg}}, \bibinfo {author} {\bibfnamefont {P.}~\bibnamefont {Olsson}},
  \ and\ \bibinfo {author} {\bibfnamefont {S.}~\bibnamefont {Teitel}},\ }\href
  {\doibase 10.1103/PhysRevLett.112.208303} {\bibfield  {journal} {\bibinfo
  {journal} {Phys. Rev. Lett.}\ }\textbf {\bibinfo {volume} {112}},\ \bibinfo
  {pages} {208303} (\bibinfo {year} {2014})}\BibitemShut {NoStop}%
\bibitem [{\citenamefont {Olsson}(2016)}]{Olsson:jam-vhist}%
  \BibitemOpen
  \bibfield  {author} {\bibinfo {author} {\bibfnamefont {P.}~\bibnamefont
  {Olsson}},\ }\href {\doibase 10.1103/PhysRevE.93.042614} {\bibfield
  {journal} {\bibinfo  {journal} {Phys. Rev. E}\ }\textbf {\bibinfo {volume}
  {93}},\ \bibinfo {pages} {042614} (\bibinfo {year} {2016})}\BibitemShut
  {NoStop}%
\end{thebibliography}%

\end{document}